%% file: main.tex
\documentclass[11pt]{article}
\usepackage{amssymb}
\usepackage{amsmath}
\usepackage{multirow}
\usepackage{arydshln}
\usepackage[toc,page]{appendix}
\usepackage{subcaption}
\usepackage{wrapfig}
\usepackage[normalem]{ulem}
\usepackage{longtable}
\usepackage{placeins}
\usepackage{etoolbox}
\usepackage{xspace}

\usepackage{graphicx}
\usepackage[margin=1.25in]{geometry}
\usepackage[usenames,dvipsnames]{color}
\usepackage{cite}
\usepackage{multirow}

\usepackage{booktabs} % for much better looking tables
\patchcmd{\thebibliography}{\section*{\refname}}{}{}{}
\usepackage[final]{pdfpages}
\usepackage{url}
\usepackage[colorlinks = true,
            linkcolor = blue,
            urlcolor  = blue,
            citecolor = blue,
            anchorcolor = blue]{hyperref}

\newcommand\pubnumber{Transcendental Preprint }
\newcommand\pubdate{\today}

\def\Title#1{\begin{center} {\LARGE #1 } \end{center}}

\newcommand{\AuthorAddress}[2]{\begin{center}{ \sc{#1}\\ \it{#2}} \end{center}}

\newcommand\pubblock{\rightline{\begin{tabular}{l} \pubnumber\\
         \pubdate \end{tabular}}}
\newenvironment{Abstract}{\begin{quotation} \begin{center}
                       ABSTRACT
     \end{center}\bigskip  }{\end{quotation}}

\def\Acknowledgements{\bigskip  \bigskip \begin{center} \begin{large}
             \bf ACKNOWLEDGEMENTS \end{large}\end{center}}
\input workshopsymbols.tex

\newcommand\snowmass{\begin{center}\rule[-0.2in]{\hsize}{0.01in}\\\rule{\hsize}{0.01in}\\
\vskip 0.1in Submitted to the  Proceedings of the US Community Study\\ 
on the Future of Particle Physics (Snowmass 2021)\\ 
\rule{\hsize}{0.01in}\\\rule[+0.2in]{\hsize}{0.01in} \end{center}}

\begin{document}

\pubblock

\Title{New directions for surrogate models and differentiable programming for High Energy Physics detector simulation}

\bigskip 
\AuthorAddress{Andreas Adelmann}{Paul Scherrer Institute, 5232 Villigen PSI, Switzerland}
\AuthorAddress{Walter Hopkins, Evangelos Kourlitis}{Argonne National Laboratory, Lemont, IL 60439, USA}
\AuthorAddress{Michael Kagan}{ SLAC National Accelerator Laboratory, Menlo Park, CA 94025, USA}
\AuthorAddress{Gregor Kasieczka}{Institut f{\"u}r Experimentalphysik, Universit{\"a}t Hamburg, Germany}
\AuthorAddress{Claudius Krause, David Shih}{NHETC, Department of Physics \& Astronomy, Rutgers University, Piscataway, NJ 08854, USA}
\AuthorAddress{Vinicius Mikuni, Benjamin Nachman}{Lawrence Berkeley National Laboratory, Berkeley, CA 94720, USA}
\AuthorAddress{Kevin Pedro}{Fermi National Accelerator Laboratory, Batavia, IL 60510, USA}
\AuthorAddress{Daniel Winklehner}{Massachusetts Institute of Technology,Cambridge, MA 02139, USA}

\medskip

 \begin{Abstract}
\noindent The computational cost for high energy physics detector simulation in future experimental facilities is going to exceed the current available resources. To overcome this challenge, new ideas on surrogate models using machine learning methods are being explored to replace computationally expensive components. Additionally, differentiable programming has been proposed as a complementary approach, providing controllable and scalable simulation routines. In this document, new and ongoing efforts for surrogate models and differential programming applied to detector simulation are discussed in the context of the 2021 Particle Physics Community Planning Exercise (`Snowmass').
\end{Abstract}

\snowmass

\def\thefootnote{\fnsymbol{footnote}}
\setcounter{footnote}{0}
%

%Would it make sense to add a brief executive summary with the main actionable recommendations?  More on this later.

\section{Introduction}

Experiments in high energy physics (HEP) rely heavily on simulation for a wide array of tasks, including data selection, statistical inference, and design optimization for new experiments. On the other hand, the computational demands for simulation of current and next generation HEP experiments have inspired investigation of surrogates, or approximations of the detector simulation, using deep generative models to decrease simulation time while maintaining fidelity. Usually, the most computationally intensive step of the simulation is the modeling of the detector response. Interactions between particles and the detector material are simulated in large experimental collaborations such as ATLAS \cite{Aad:2008zzm} and CMS \cite{Chatrchyan:2008aa} using the \geant~\cite{geant4,geant4-add1, geant4-add2} software package. While full simulation ensures high fidelity samples, the computational cost becomes prohibitive as many billions of simulated events are required to describe different Standard Model  and Beyond the Standard Model  processes. For comparison, detector simulation in the ATLAS and CMS experiments consumed 40\% of the grid central processing unit (CPU) during Run 2 of the LHC experiment \cite{HEPSoftwareFoundation:2018fmg,ATLAS:2021pzo}, and the expected CPU time needed to simulate an event increasing by a factor of three \cite{Pedro:2018jqu} or more after the HL-LHC upgrade in the upcoming years.

%Should we add a brief comment (maybe as a footnote) that there is much work on classical attempts at fast simulations?  We should briefly say why deep generative models are needed.

Generative models leveraging recent advancements in machine learning (ML) are able to build surrogate models capable of generating high fidelity samples with reduced computational cost. Common software frameworks for ML research, like  \texttt{TensorFlow}~\cite{tensorflow2015-whitepaper}, \texttt{JAX}~\cite{jax2018github}, or \texttt{PyTorch}~\cite{NEURIPS2019_9015} benefit from strong community support and highly efficient implementations on hardware accelerators, such as Graphics Processing Units (GPUs). This flexibility is easily ported to experimental facilities and lowers the barrier of entry for software development, support, and maintenance. These improvements have the possibility to accelerate the comparison between measurements and theoretical predictions while decreasing the need for methods such as unfolding \cite{Tikhonov:1963,DAgostini:1994fjx,Hocker:1995kb} once an efficient detector simulation is available. 
% Additional applications for fast surrogate models include optimization and design of particle accelerators \cite{edelen:ml2,Koser:ML1,van_der_veken:ml1}, real-time feedback during commissioning and tuning of an accelerator facility \cite{Koser:ML1, van_der_veken:ml1}, and uncertainty quantification of simulated parameters \cite{adelmann:surrogate1,Winklehner:2021qzp}. %However, what if we build our model in JAX and then it stops being supported?  Are we in trouble?  Do we need HEP community members to be paid to support these tools to ensure usability and continuity for HEP? (I think the answer is yes!)

% Generalizing such approaches, the emerging new paradigm of differentiable programming (DP) can provide powerful new directions in simulator modeling that (a) increase the amount of information one can extract from simulated data, thus potentially reducing the required scale of simulated datasets used for downstream tasks, (b) incorporate known physics knowledge, which is critical for developing more robust, interpretable, and generalizable domain-aware scientific ML~\cite{osti_1478744}, and (c) can be integrated in ML inference pipelines for building physics-informed models, enabling the development of new inference methods.

Traditional simulation routines can be improved in multiple ways, leveraging modern software frameworks and hardware accelerators, such as GPUs \cite{Carrazza:2020rdn,Carrazza:2021gpx,10.3389/fdata.2021.665783}. Alternatively, differentiable programming (DP) software can also enable GPU support to traditional simulation routines. Differentiable programs track gradients with respect to simulation parameters or input variables at each step of the simulation program. While DP is not required for an algorithm to benefit from modern hardware acceleration (and not all DP frameworks are inherently GPU-compatible), differentiable programs provide additional advantages. For example, optimization of simulation inputs can be directly inferred from experimental data by finding the simulation parameters that jointly minimize the difference between synthetic and experimental data. The optimization step is performed by propagating the gradients back through the simulation chain, thus reducing the required scale of simulated datasets used for alternative setups\cite{Brehmer:2018hga}. DP also opens new directions for simulation modeling that incorporate physics knowledge, which is critical for developing more robust, interpretable, and generalizable domain-aware scientific ML~\cite{osti_1478744}.

Additional usage of surrogate modeling and differentiable programming for HEP in the context of Snowmass are covered in detail in the Snowmass'21 LOIs \cite{LOI_ML1, LOI_ML2},
the upcoming Snowmass'21 whitepaper by the Beam and Accelerator Modeling Interest 
Group (BAMIG) \cite{vay:snowmass21}, two recent ICFA newsletters 
\cite{vay_modeling_2021, sagan_simulations_2021}, and the MODE collaboration~\cite{MODE:WP}.

%In this document, ongoing ideas proposed to decrease the computational cost of detector simulation in HEP are discussed. Those include surrogate models and differentiable programming for detector simulation. 
This document is divided as follows. Sections~\ref{sec:works} and \ref{sec:diffprog} present brief introductions to surrogate models based on ML techniques and differentiable programming. In Secs.~\ref{sec:contrib:EK_WH}, \ref{sec:contrib:CK}, and \ref{sec:contrib:MK}, several ongoing projects are described and discussed. These examples are not meant to be comprehensive, but instead illustrative of the scope of research in this area. Finally, in Sec.~\ref{sec:future}, future directions and synergies in the short and long term future are explored.

\section{Surrogate Models}\label{sec:works}

% taken from the footnote previously in introduction
The existing landscape of detector simulation consists of two primary approaches. The first is the accurate, but computationally intensive ``full simulation'' using \geant. The second is typically called ``fast simulation'' and may be considered a classical version of a surrogate model. Decreased simulation time is achieved by replacing computationally intensive parts of the simulation with simplified detector assumptions, resulting in speed improvements of more than 100 times compared to the full simulation. The resulting simulation, however, is less realistic and may be unsuitable for physics measurements that rely on detailed detector effects. Within the category of classical fast simulation, there are experiment-specific solutions~\cite{ATLAS:2010arf,ATLAS:2021pzo,Sekmen:2016iql} and the ultra-fast generic simulation \delphes~\cite{deFavereau:2013fsa}.

\begin{wrapfigure}{l}{0.5\textwidth}
    \centering
    \includegraphics[width=0.4\textwidth]{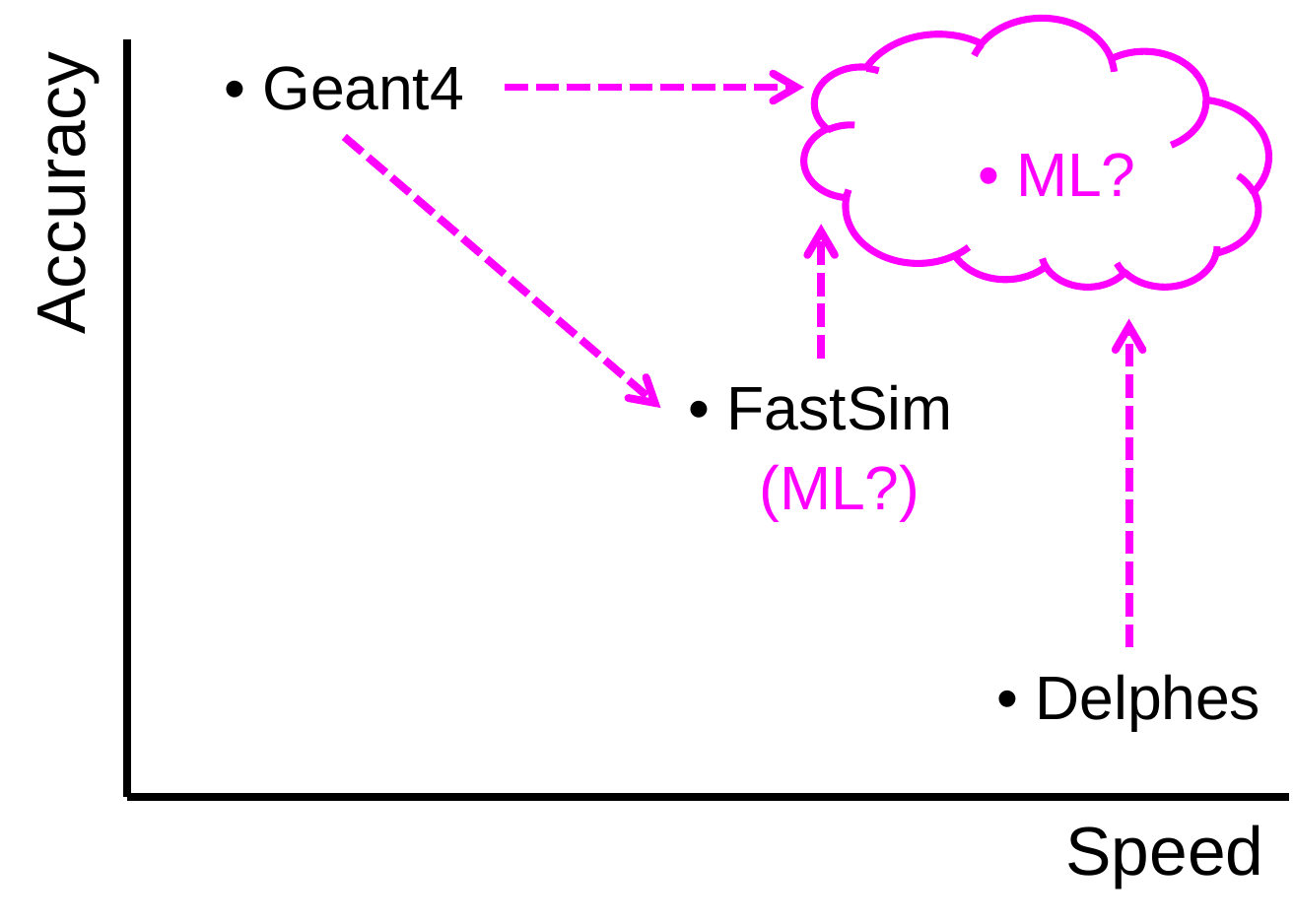}
    \caption{Depiction of different ways to incorporate ML in detector simulation workflows.} 
    \label{fig:ml4sim_diagram}
\end{wrapfigure}

Figure~\ref{fig:ml4sim_diagram} shows the different ways to introduce ML to this landscape: by replacing or augmenting part or all of \geant, or part or all of a classical fast simulation. Each option has a different goal: increasing speed while preserving accuracy, or preserving speed while increasing accuracy, respectively. ML could also be used to create a faster but less accurate simulation, similar to existing classical fast simulations.
Alternatively, different ML surrogate models approaches may be classified based on what input data they require to produce simulated events. This leads to two categories: 1. fully generative models that entirely replace classical simulation engines, taking generated particle data or random noise as input; and 2. refinement techniques that are applied during or after the event simulation step, taking lower-quality simulated events as input. Popular deep learning architectures for fully generative models are divided into three main categories including generative adversarial networks (GANs) \cite{Paganini:2017hrr,Paganini:2017dwg,deOliveira:2017rwa,DiSipio:2019imz,Farrell:2019fsm,Erdmann:2018kuh,Erdmann:2018jxd,Deja:2019vcv,Musella:2018rdi,Vallecorsa:2018zco,Carminati:2018khv,ATL-SOFT-PUB-2018-001,Chekalina:2018hxi,SHiP:2019gcl,Martinez:2019jlu,Butter:2020tvl,2009.03796,Kansal:2020svm,Maevskiy:2020ank,Choi:2021sku,Rehm:2021zow,Rehm:2021zoz,Kansal:2021cqp,Khattak:2021ndw,Bravo-Prieto:2021ehz,ATLAS:2021pzo}, variational autoencoders (VAEs) \cite{ATL-SOFT-PUB-2018-001,Buhmann:2021lxj,Hariri:2021clz,Mu:2021nno,Orzari:2021suh}, and normalizing flows \cite{Lu:2020npg,Krause:2021ilc,Krause:2021wez, Butter:2021csz}. Refinement techniques may be based on classification \cite{2009.03796,Winterhalder:2021ave} or regression \cite{Chen:2020uds,banerjee2022denoising}. The generative models apply a stochastic approach, while the refinement techniques are usually deterministic.

Initially proposed in \cite{NIPS2014_5ca3e9b1}, GANs are trained following a minimax game:
\begin{equation}
    \min_{G}\max_{D}V(D,G) =  \mathbb{E}_{x\sim p_x(x)}[\log D(x)] + \mathbb{E}_{z\sim p_z(z)}[\log(1-D(G(z)))]\,,     
\end{equation}
where a generator network $G$ is tasked to generate new samples from a noise distribution $p_z(z)$ while the discriminator network $D$ judges the quality of the generated samples by comparing with target events sampled from $p_x(x)$. The adversarial loss function can lead to unstable training, often requiring additional fine tuning to achieve realistic results. An alternative to the loss function was proposed in \cite{arjovsky2017wasserstein} named Wasserstein GAN (WGAN). In the WGAN framework, the discriminator network is replaced with a critic network that uses the Wasserstein distance between generated and data samples as a metric to be minimized during training. There are many GAN variations that go beyond the vanilla and WGAN approaches.

Autoencoders are composed of two components: an encoder that compresses a set of input features into a smaller latent space, and a decoder that uses the information in that latent space to attempt to reconstruct the input features.  VAEs combine autoencoders with a tractable latent space to generate new and realistic samples. Even though the probability density of the data is not tractable, VAEs minimize the evidence lower bound loss:
\begin{equation}
    \mathcal{L}_{\mathrm{VAE}} = -\mathbb{E}_{z\sim q(z|x)}[\log p_x(x|z)] + \mathcal{D}_{KL}(q(z|x)||p_z(z)).
\end{equation}
The approximate posterior probability density $q(z|x)$ is enforced to be a tractable  distribution through the Kullback–Leibler  divergence term $\mathcal{D}_{KL}$. The reconstruction loss $\log p_x(x|z)$ is often defined as the mean squared error loss, in case of continuous distributions, or the cross-entropy loss, in case of discrete distributions.

As an alternative approach to handling data with an intractable probability distribution, normalizing flows~\cite{2015arXiv150505770J,2019arXiv190809257K,2019arXiv191202762P} define a bijective transformation between a tractable base distribution, such as a normal or uniform distribution, to the data using the transformation of variables:
\begin{equation}
    \log{p_{x}(x)} =  \log{p_{z}(z)} -\log\mathrm{det}\left| \frac{\partial f(z)}{\partial z}\right|,
    \label{eq:flow}
\end{equation}
with terms $\frac{\partial f(z)}{\partial z}$ representing the Jacobian matrix of the transformation $f$. The loss function to be minimized is then defined as $-\log{p_{x}(x)}$, which is equivalent to minimizing the KL divergence between the transformed tractable base distribution and the data distribution.

\begin{wrapfigure}{l}{0.5\textwidth}
    \centering
    \includegraphics[width=0.5\textwidth]{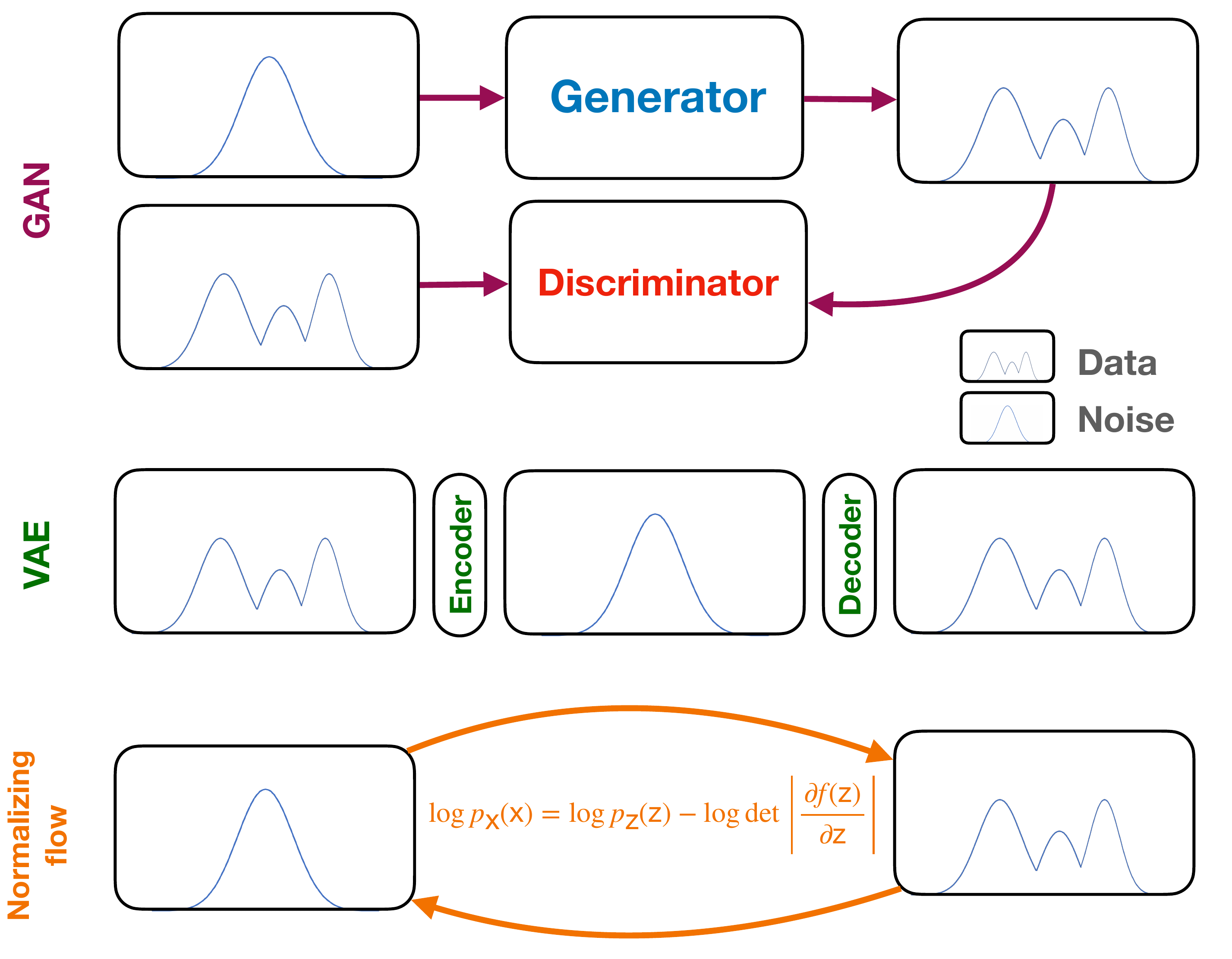}
    \caption{Summary of different machine learning methods used for generative models.} 
    \label{fig:gan_chart}
\end{wrapfigure}

The general calculation of the determinant in Eq.~\ref{eq:flow} has $\mathcal{O}(D^3)$ computational cost. This limitation is mitigated by restricting the bijective transformation $f$ to the family of functions with triangular Jacobian matrix, bringing the computational complexity down to $\mathcal{O}(D)$. 
There exist two main architectures to ensure a triangular Jacobian: bipartite~\cite{2016arXiv160508803D} flows based on so-called coupling layers or autoregressive~\cite{2017arXiv170507057P,2016arXiv160604934K} flows based on masked neural networks (NNs)~\cite{2015arXiv150203509G}. Practically speaking, their main difference is the speed in which both directions of the bijector can be evaluated. In bipartite flows accessing the log-likelihood of data and sampling is equally fast. Autoregressive flows have a fast and a slow direction. Masked Autoregressive Flows (MAFs)~\cite{2017arXiv170507057P} are fast in density estimation, but a factor $D$, given by the dimension of data space, slower in sampling. Inverse Autoregressive Flows (IAFs)~\cite{2016arXiv160604934K} are fast in sampling, and a factor $D$ slower in estimating the density of data points. See~\cite{2019arXiv190809257K,2019arXiv191202762P} for more details on normalizing flows.
An application of density estimation for calorimeter simulation is described in Sec.~\ref{sec:contrib:CK}. A simplified depiction of the different generative strategies is shown in Fig.~\ref{fig:gan_chart}.

A natural question about surrogate models is to what extent the generated samples increase the statistical power with respect to the training data. At its core, the benefit from deep generative models comes from their ability to interpolate in high dimensions. One source of statistical amplification from the training dataset is from combinatorics - there are combinatorially many ways to attach showers to $N$ particles in an event and deep generative models can naturally interpolate from the training dataset to have the correct kinematic properties.  Interpolation also can result in improved statistical precision from the smoothness properties of neural networks (a form of `inductive bias')~\cite{2008.06545,Bieringer:2022cbs}.

Further refinements to generated samples can be derived to improve generation quality. Those corrections can be coupled either to ML-based surrogate models or to classical fast simulation routines. Advantages of refinement in the former case include that the training is often more stable than for the original generative model, and correspondingly, the generative model may not need to match the precision of \geant. Alternatively, replacing the generative model with a classical simulator and relying on ML only for refinement may decrease the probability of unphysical output and provide better extrapolation beyond the training data. Some example applications of this type are discussed in Sec.~\ref{sec:contrib:EK_WH}.

\section{Differentiable Programming}\label{sec:diffprog}

An alternative but complementary direction for surrogates lies in recent advancements in differentiable programming (DP).  In DP, software is written in, or transformed into, differentiable code via the use of automatic differentiation (AD)~\cite{autodiff}, an algorithmic way to efficiently evaluate derivatives of computer programs. When software is written in DP frameworks, access to the dependence of predictions on inputs is enabled through gradients. 

\begin{wrapfigure}{r}{0.5\textwidth}
    \centering
    \includegraphics[width=0.5\textwidth]{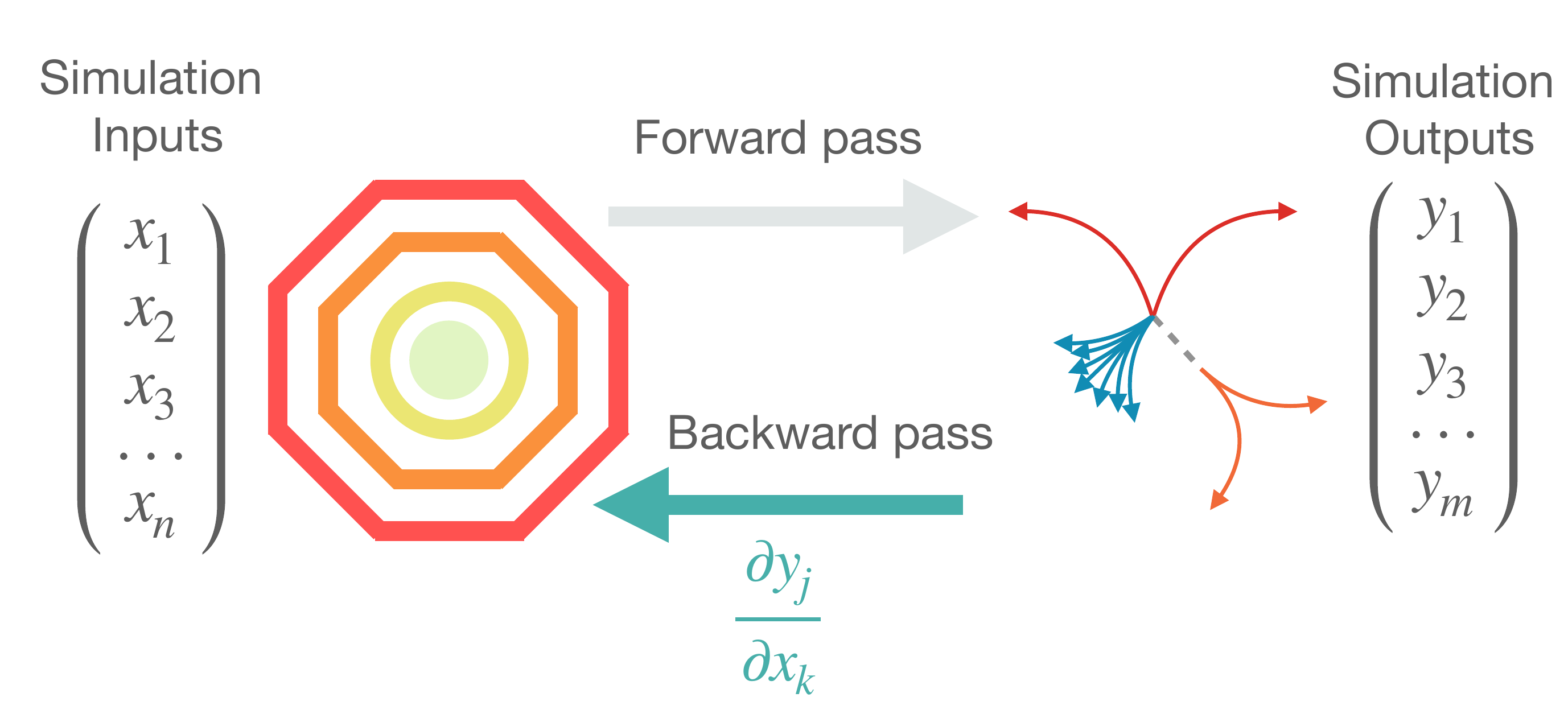}
    \caption{An illustration of differentiable programming for detector simulation.} 
    \label{fig:diff_chart}
\end{wrapfigure}

These gradients are a significant addition to the information typically provided by simulators and crucially can be used in downstream modeling and inference tasks. This approach is flexible and optimizable; differentiable HEP software and ML tools can be mixed, for instance to use ML surrogates of non-differentiable computations, and can be jointly optimized to improve speed and prediction accuracy. When developed with DP, HEP simulation tools, and the physics knowledge they encode, can be used as physics prediction engines directly within ML pipelines for developing physics-informed ML tools. An illustration of differentiable programming is shown in Fig.~\ref{fig:diff_chart}. We note that deep generative models are a type of differentiable detector simulation, since gradients are readily available for neural networks. More details on differentiable programming for detector simulation are given in Sec.~\ref{sec:contrib:MK}.

\section{ML-based Correction to Accelerate \geant\ Calorimeter Simulations}\label{sec:contrib:EK_WH}

% The \geant~\cite{geant4,geant4-add1, geant4-add2} package provides the full interaction between particles and the experimental apparatus. 
In full simulation routines, particles can be fully tracked using the complete underlying physics knowledge (FullSim) or approximate parametrizations can be used to simplify and accelerate the process (FastSim). Although future experiments plan to be heavily based on FastSim methods, the usage of FullSim is still imperative \cite{Calafiura:2729668} (including FastSim tuning). % The latter, among others, for calibration and FastSim tuning purposes.

Focusing on the FullSim, among the most computationally demanding apparatuses to simulate are dense highly segmented particle physics detectors (e.g. calorimeters). This is because highly energetic particles produce cascades of secondary particles, resulting in an exponential number of particles with respect to the particle energy.  The actual limit on the lowest energy particle simulated is controlled by \textit{range cuts}.
% In such cases, geometrical boundaries mainly limit the simulation steps in spacetime, rather than physics interactions. This is often the case for highly segmented electromagnetic calorimeters. Detailed studies have shown that the majority of the computational load originates from low-energy photons (O(\MeV)). These are produced during the shower evolution and propagate through the detector, with few interactions due to their neutral charge. The production of secondary particles is controlled in \geant\ by energy thresholds (per particle, per material). These are adjusted by internal parameters called range cuts.
Increased range cuts correspond to increased production energy thresholds, thus reducing the number of produced secondary particles. As an immediate effect, the computational demands of the simulation are reduced. A side effect can be the reduction of the accuracy of the simulation. The extent of the inaccuracy increases as the range cut grows relative to the scale of the sensitive elements of the detector. While other parameters in \geant\ may also be varied with effects on the simulation computing time and accuracy, range cuts have been found to be the most impactful.

This section outlines an approach to accelerate the FullSim execution time.  One possibility is to use a deep generative model as a base that is then refined~\cite{2009.03796}. Another approach is applying aggressive range cuts and then correcting the reduced accuracy simulation (referred to as \textit{modified}) using ML methods. There are several complementary techniques to derive these corrections.

One method relies on event-level weights. A neural network is trained to classify the nominal versus the modified simulation and the classification score is used to calculate a multi-dimensional density ratio. This ratio is finally used to reweight the modified observables back to the nominal ones. The classification score is then used to approximate the density ratio $r(x)$. There are many ways to do this (see Ref~\cite{Nachman:2021yvi}), but the most common approach is to use the binary cross entropy loss function and then derive $r$ as:
\begin{equation} \label{eq:density_ratio}
   r(\bar{x})= \frac{\rho(y=1|x)}{1-\rho(y=1|x)} \approx \frac{p_\text{full}(x)}{p_\text{fast}(x)}\,,
\end{equation}
where $\rho$ is the classifier and $x$ is a set of observables used in the reweighting. The approximation in the above equation is a well-known result from statistics (see e.g.,~\cite{hastie01statisticallearning,sugiyama_suzuki_kanamori_2012}). Additional post-processing can improve the approximation~\cite{Cranmer:2015bka}.

%$\theta$ is the applied range cut and $\bar{x}$ the input observable vector. The nominal and alternative probability density functions are represented by $p$ and $q$, respectively. This density ratio can be used to reweight the observables of the alternative simulation back to the ones of the nominal density. Theoretical background and proof of the method are given in \cite{Cranmer:2015bka}. 

There is no unique way to pick $x$.  One possibility is to refrain from choosing any specific high-level observables and instead learn directly from the lowest level inputs (e.g. energy deposits per calorimeter cell).  An advantage of using high-level features is that it provides some regularization so that if the original model has phase space gaps in high-dimensions, there will not be infinities in the likelihood ratio~\cite{2009.03796}.

\begin{figure}[!ht]
    \centering
    \includegraphics[width=0.6\textwidth]{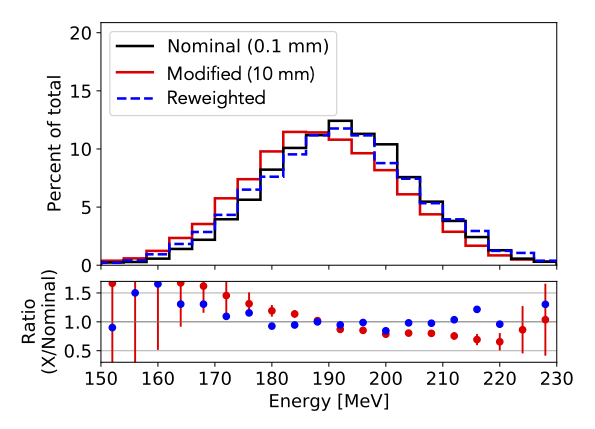}
    \caption{Comparison of the nominal, modified, and reweighted event energy deposit at the ILD barrel calorimeter induced by 10 GeV electron showers. The nominal distribution uses a 0.1 mm range cut, while the modified uses 10 mm, leading to ${\sim}15\%$ simulation CPU speedup.} 
    \label{fig:ILD_edep}
\end{figure}

Serving as a proof of concept, an example reweighting application using the lowest level inputs has been developed for the International Large Detector (ILD) electromagnetic barrel calorimeter~\cite{ILDConceptGroup:2020sfq}. The multilayer calorimeter consist of 30 layers, each one segmented in $30\times30$ cells. The data are projected into 3D images, where the color of each of the 27,000 voxels represent the energy deposit in the cell. A convolutional neural network (CNN) utilizing 3D convolution operations is trained to discriminate nominal from modified simulation events in order to approximate the ratio of Equation~\ref{eq:density_ratio}. Preliminary results shown in Figure~\ref{fig:ILD_edep} showcase the improvement of the reweighted modified simulation, resembling the high accuracy nominal \geant\ simulation. The tradeoff of the correction via reweighting is the statistical dilution of the simulation sample~\cite{2009.03796}.

%In order to retain the generality of the method, we refrain from choosing any specific high-level observable to bin and extract the aforementioned density ratio. Instead, we use as inputs to the classifier the energy deposits per calorimeter cell. The extracted multi-dimensional weight can then applied to correct any observable.

\begin{figure}[!hb]
  \centering
  \includegraphics[width=0.49\textwidth]{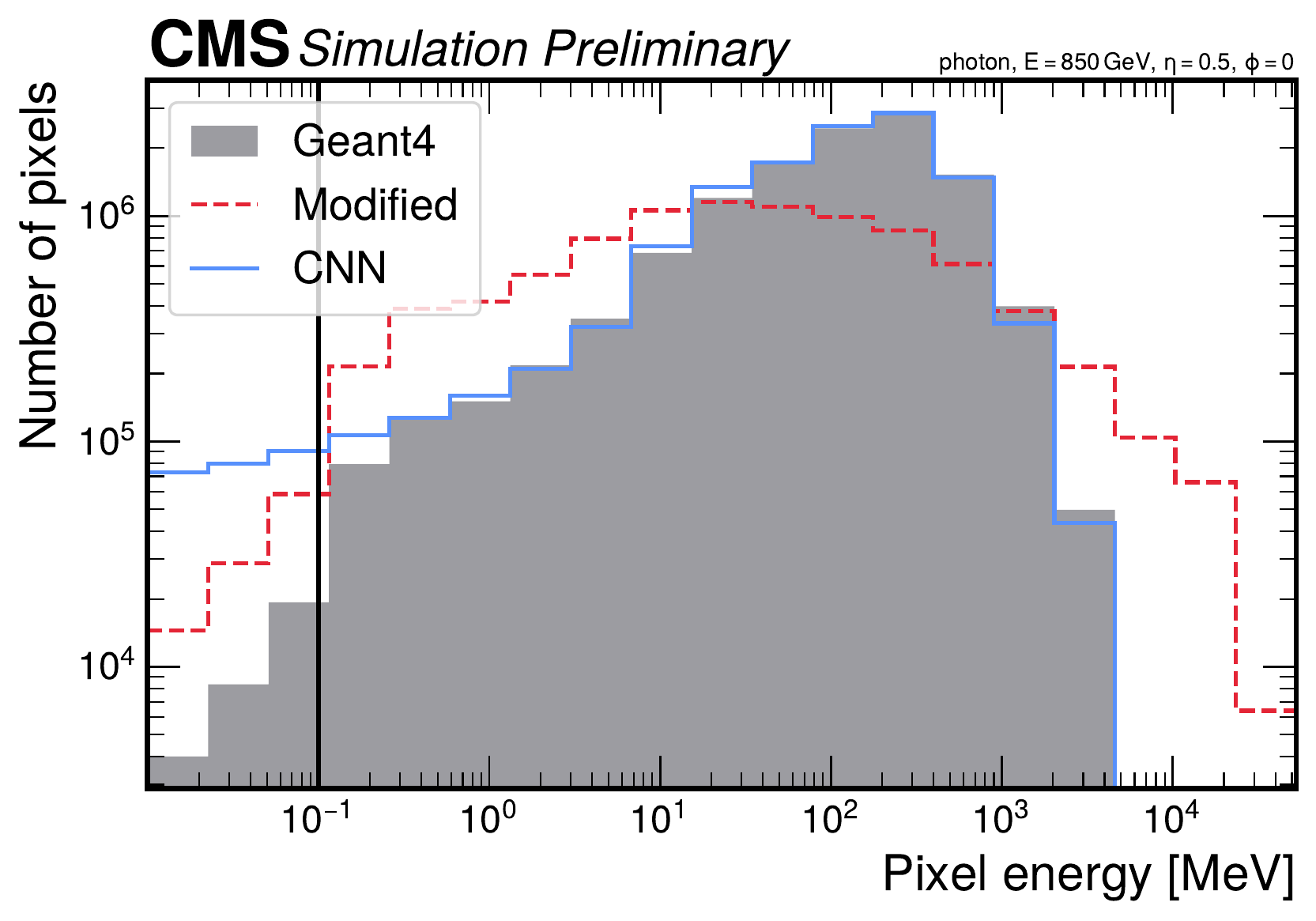}
  \includegraphics[width=0.49\textwidth]{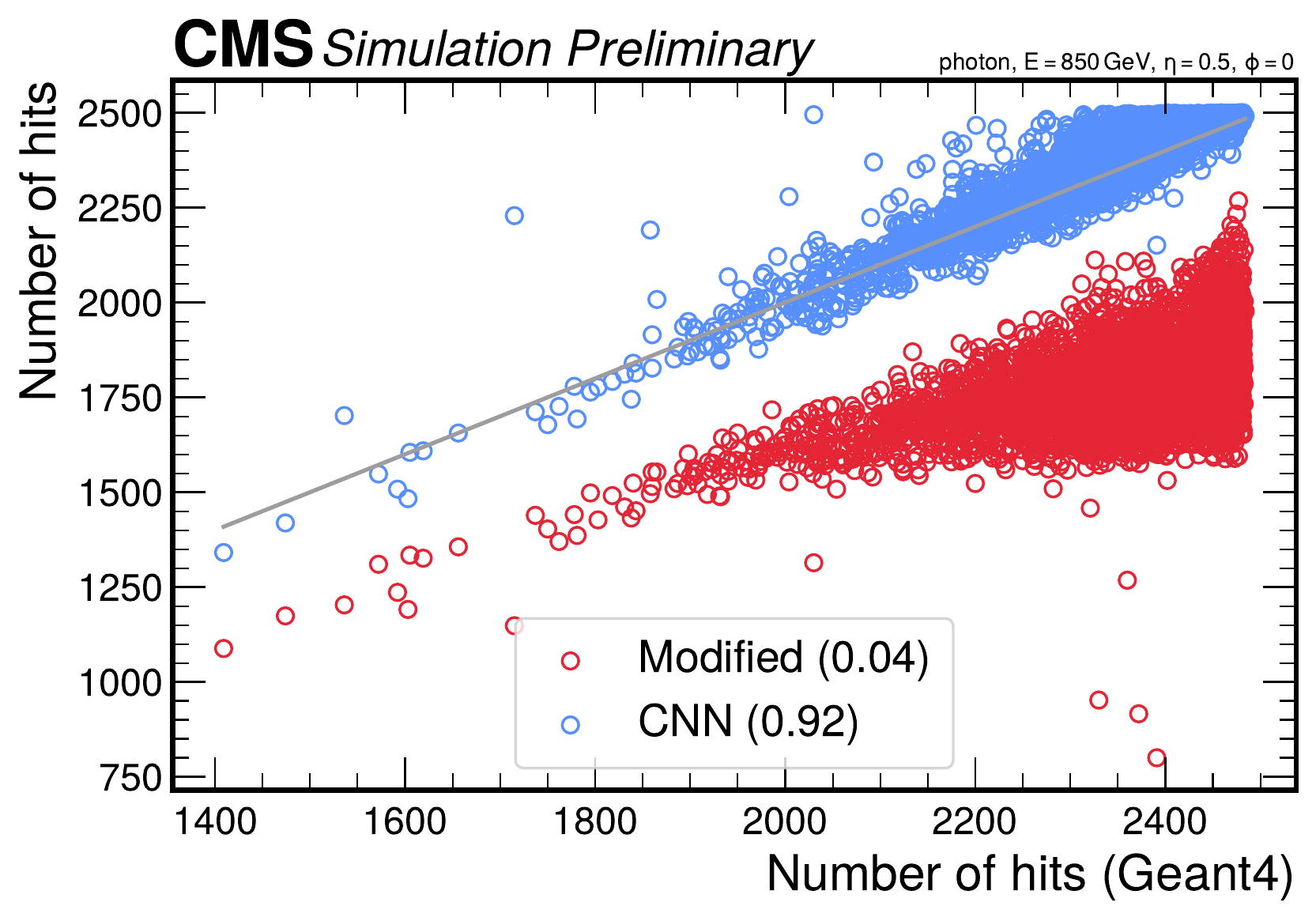}
  \caption{Left: A comparison of the per-pixel energy distribution for the modified simulation, the CNN output, and \geant. Right: Per-event comparisons of the number of hits, with the concordance correlation~\cite{Concordance} between \geant\ and the other simulations listed in parentheses and the gray line indicating exact agreement. These figures are reproduced from Ref.~\cite{banerjee2022denoising}.}
  \label{fig:denoising_results}
\end{figure}

An alternative method directly modifies the simulated event contents. A first proof of concept for this method is described in Ref.~\cite{banerjee2022denoising}, based on an approach used in industry to accelerate MC ray-tracing \cite{10.1145/3072959.3073708}, has recently been published. \geant\ with an increased range cut provides the modified simulation, and a CNN is used to regress the energy value of each pixel, with the detector represented as a digitized grid. Figure~\ref{fig:denoising_results} shows the promising results for photon showers in the CMS electromagnetic calorimeter. Regression approaches can also be applied to improve high-level variables, which may complement the low-level approach.

% is it also worth reproducing Fig. 3 from banerjee2022denoising, showing an example event with different range cuts?

These correction approaches can reduce the computational complexity for two reasons: 1) the absolute time reduction of the simulation from a faster surrogate (either a deep generative model or higher range cuts) can reduce the calculation time by orders of magnitude and 2) the correction may be applied in parallel to many events utilizing parallel computations in accelerator hardware, such as GPUs. The overall speedup from this approach may be limited by the throughput of the classical fast simulation engine, in this case the modified \geant. However, the potential for greater accuracy and reliability may make such a trade-off worthwhile, if the overall speedup is enough to meet the computing challenges of the HL-LHC and future colliders.

\section{Detector Simulation with Normalizing Flows}\label{sec:contrib:CK}
 Ideal surrogate models are fast and at the same time indistinguishable from the full simulation based on \geant. The latter can be tested by training a neural network classifier on ``real'' (based on \geant) and ``fake'' (generated from the surrogate) samples~\cite{2016arXiv161006545L}. Previous surrogate models, based simple on GANs or VAEs, have failed such a test and yielded samples that were separable to nearly 100\%.
 Normalizing Flows (NFs) provide an alternative approach to generative modeling, since they learn the likelihood of the data explicitly, in contrast to GANs and VAEs that only implicitly learn the data distribution. Maximizing the log-likelihood of the training data directly is more stable and not prone to mode-collapse. 
 In addition, picking the model with the lowest validation loss seems to be an effective model selection strategy, which is a challenge for deep generative adversarial models.
 %In addition, it allows a model selection that objectively captures the entire feature space with all its correlations.
 
 The proof of concept of this approach is given in~\cite{Krause:2021ilc,Krause:2021wez} (called~\textsc{CaloFlow}), based on the same detector geometry that was studied in~\cite{Paganini:2017hrr,Paganini:2017dwg}. This geometry is a simplified version of the ATLAS electromagnetic (ECAL) calorimeter, consisting of 3 layers with 288, 144, and 72 voxel, respectively. A new instance of~\textsc{CaloFlow} was trained for each particle type ($e^{+}$, $\gamma$, $\pi^{+}$).  In~\textsc{CaloFlow}, the data likelihood is learned in two steps, with two separate NFs. The first step only learns how the total deposited energy is distributed across the three calorimeter layers, conditioned on the incident energy, $p_1(E_i|E_{\text{inc}})$, with $E_{\text{tot}} = \sum_{i} E_{i}$. \textsc{CaloFlow} uses a MAF for flow 1. The second step learns the normalized shower shape, i.e.~how the energy deposited in each layer is distributed into the voxels, conditioned on the energy deposition of each layer and the incident energy, $p_2(\mathcal{I}| E_i,E_{\text{inc}})$. Both autoregressive architectures, MAFs and IAFs, have been applied to this step in~\cite{Krause:2021ilc} and~\cite{Krause:2021wez} respectively. However, the high dimensionality of the voxel space made a training based on the log-likelihood prohibitive for the IAF. Instead, the flow of~\cite{Krause:2021wez} was trained using probability density distillation, a method originally developed for speech synthesis in~\cite{2017arXiv171110433V}. In generation, one first samples $E_{i}$ from flow 1. These energies are then given to flow 2 to generate the showers. After shower generation, the resulting showers are renormalized to have the energies according to the $E_{i}$ of flow 1.
 
\begin{figure}[!ht]
  \centering
  \includegraphics[width=0.75\textwidth]{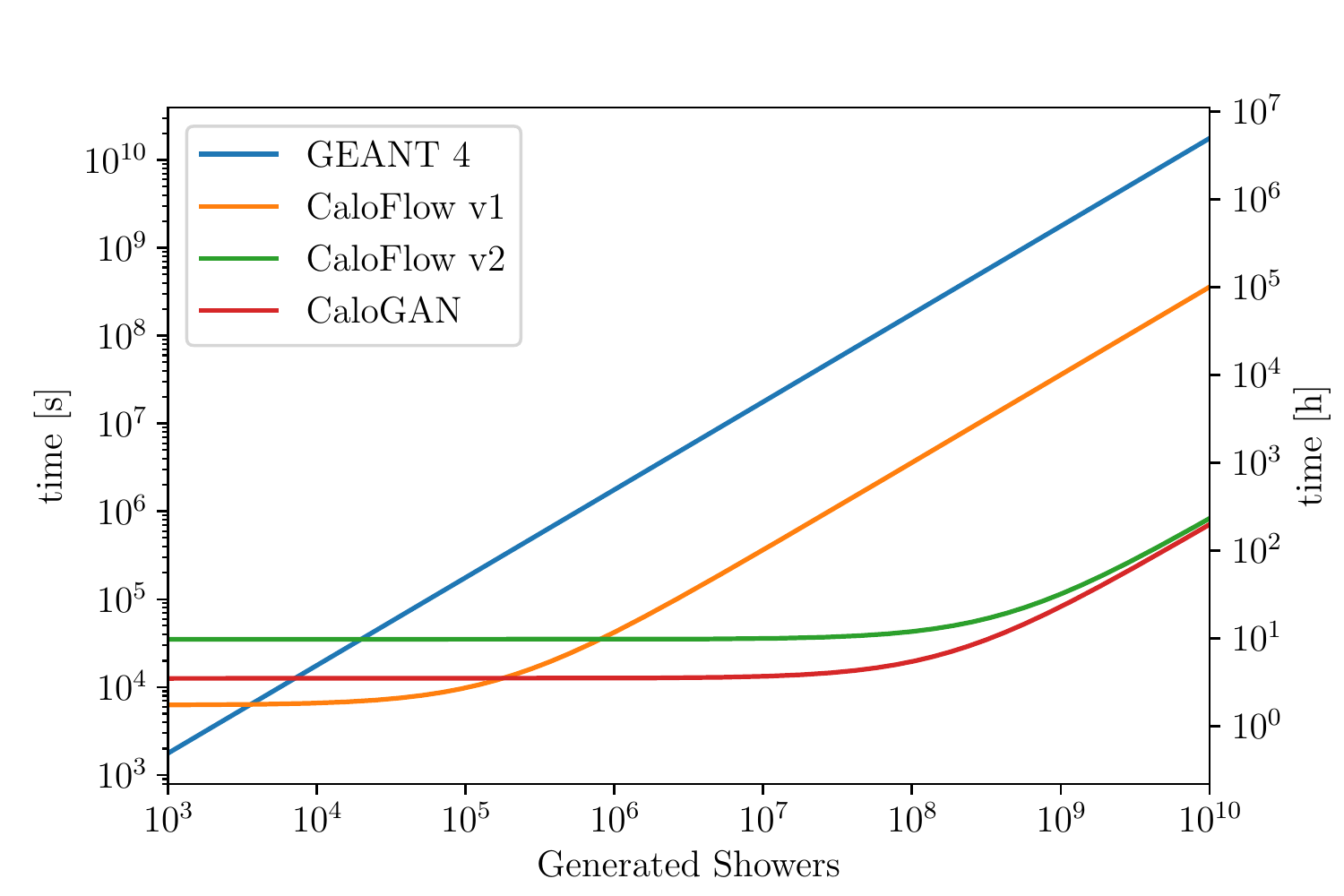}
  \caption{Comparison of shower generation times of \geant, CaloGAN~\cite{Paganini:2017hrr,Paganini:2017dwg}, CaloFlow v1~\cite{Krause:2021ilc}, and CaloFlow v2~\cite{Krause:2021wez}.}
  \label{fig:caloflow.timing}
\end{figure}

\begin{figure}[!ht]
    \centering
    \includegraphics[width=\textwidth, trim=50 0 75 50, clip]{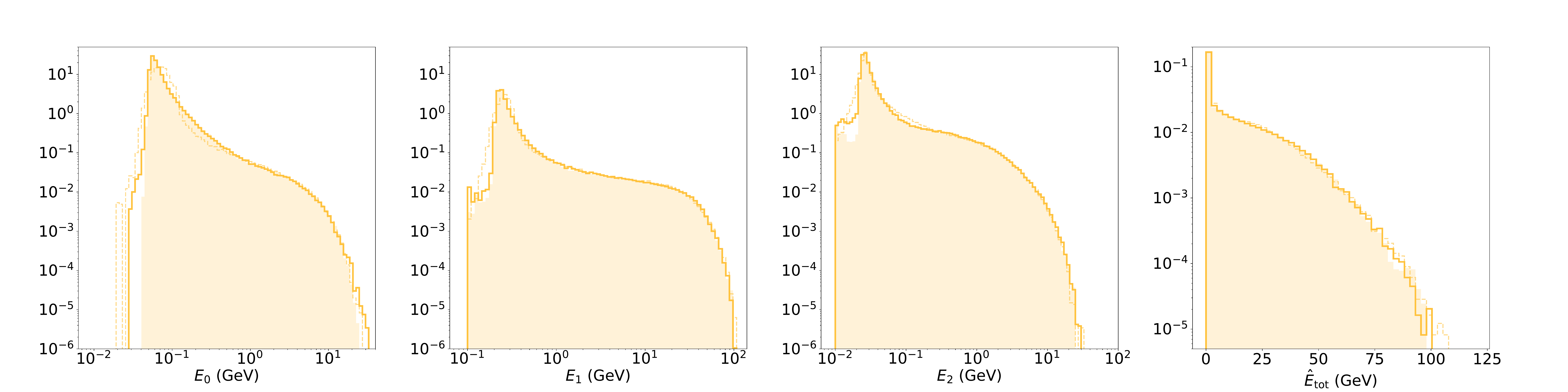}
    \includegraphics[width=\textwidth, trim=75 700 100 50, clip]{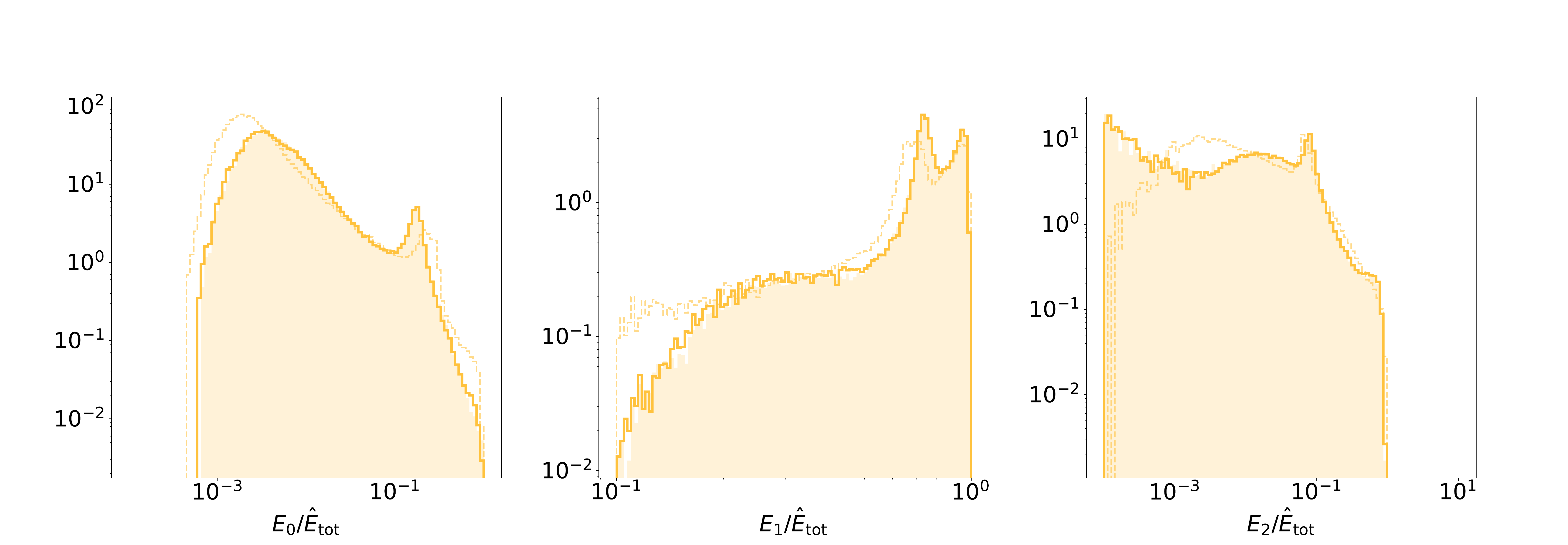}
    \includegraphics[width=0.75\textwidth]{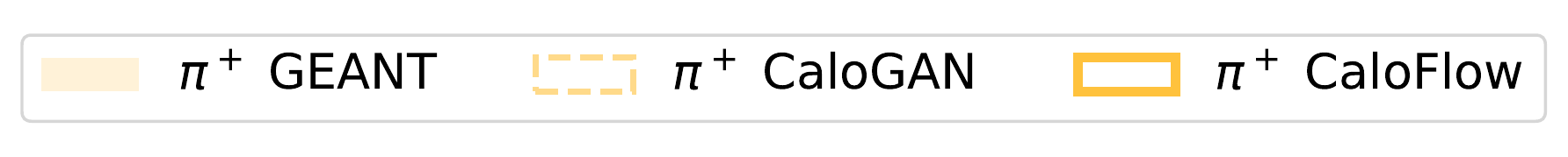}
    \caption{Distributions of energies in the 3 Calorimeter layers and total deposited energy (top) and ratio of layer energies to total deposited energy (bottom) for incident $\pi^+$ particles, comparing \geant\ to CaloGAN~\cite{Paganini:2017hrr,Paganini:2017dwg} to CaloFlow~\cite{Krause:2021ilc,Krause:2021wez}.} 
    \label{fig:caloflow.histos.piplus}
\end{figure}

 Table~\ref{tab:CaloFlow.classifier} shows the main results of that approach, given by the training of a binary neural classifier. While this GAN-based model yields samples that are distinguishable from the \geant\ samples, the NF-based model has a much higher fidelity and can fool the classifier much more often. \footnote{It is also possible that more recent, state-of-the-art GANs or VAEs perform better on this dataset, which is an interesting topic for future studies.} The generation of samples, especially with \textsc{CaloFlow} v2, is as fast as the GAN. Differences in training time become irrelevant once more than $10^9$ showers are generated, see Fig.~\ref{fig:caloflow.timing}. Figure~\ref{fig:caloflow.histos.piplus} shows some example distributions for $\pi^+$ showers, comparing \geant\ to CaloGAN~\cite{Paganini:2017hrr,Paganini:2017dwg} to CaloFlow~\cite{Krause:2021ilc,Krause:2021wez}.

 \renewcommand{\arraystretch}{1.5}
\begin{table*}[!ht]
\caption{AUC and JSD metrics for the classification of \geant\ vs \textsc{CaloGAN}, \textsc{CaloFlow} v1, and \textsc{CaloFlow} v2 showers. Classifiers were trained on each particle type ($e^{+}$, $\gamma$, $\pi^{+}$) separately. All entries show mean and standard deviation of 10 runs and are rounded to 3 digits (lower numbers are better). Taken from~\cite{Krause:2021wez}. } 
\label{tab:CaloFlow.classifier}
%\hspace*{-2em}
\begin{tabular}{|c|c|c|c|c|}
  \hline
  \multicolumn{2}{|c|}{} & \multicolumn{3}{c|}{DNN-based classifier} \\
  \cline{3-5}
  \multicolumn{2}{|c|}{AUC / JSD} & \multicolumn{3}{c|}{\geant\ vs.} \\
  \multicolumn{2}{|c|}{} & \textsc{CaloGAN} & \textsc{CaloFlow} v1 & \textsc{CaloFlow} v2 \\
  \hline
  \multirow{2}{*}{$e^{+}$}&unnormalized & \; 1.000(0) / 0.995(1) \; & 0.859(10) / 0.365(14)  &0.786(7) / 0.201(11) \\
  \cline{2-5}
  &  normalized & \; 1.000(0) / 0.997(0) \; & 0.870(2) / 0.378(5) & 0.824(4) / 0.257(8) \\
  \hline
  \multirow{2}{*}{$\gamma$}&unnormalized & \; 1.000(0) / 0.998(0) \; &  0.756(48) / 0.174(68) & 0.758(14) / 0.162(18) \\
  \cline{2-5}
  &  normalized & \; 1.000(0) / 0.994(1) \; & 0.796(2) / 0.216(4) & 0.760(3) / 0.158(4) \\ 
  \hline
  \multirow{2}{*}{$\pi^{+}$}&unnormalized & \; 1.000(0) / 0.993(0) \; & 0.649(3) / 0.060(2) & 0.729(2) / 0.144(3) \\
  \cline{2-5}
  &  normalized & \; 1.000(0) / 0.997(1) \; & 0.755(3) / 0.153(3) & 0.807(1) / 0.230(3) \\
  \hline
\end{tabular}
\end{table*}
\renewcommand{\arraystretch}{1}

In order for normalizing flow-based models to be used by the experimental collaborations, they have to prove their performance in more realistic setups as well. These will either have a higher number of voxels (like the ILD or CMS high granularity~\cite{CMS:2017jpq} calorimeters) and/or a conditioning on incident angle and/or position. Additional studies on the specific NF architecture, such as autoregressive vs. bipartite flows, convolutional and other types of networks to give parameters of the transformation, or new bijective transformations might reveal more efficient (in terms of memory usage and/or sampling time) setups. Novel setups might also circumvent the 2-step approach of \textsc{CaloFlow}.

\section{Simulation of increasingly complex detectors }\label{sec:contrib:GK}

A number of challenges is encountered when moving from simplified detectors towards realistic simulations of energy deposits in modern calorimeters.

A primary issue lies in the substantial number of hits that need to be simulated. For example, the planned CMS High-Granularity Calorimeter (HGCAL)~\cite{CERN-LHCC-2017-023} will have $\approx$ 6 million individual read-out channels with a similar order of magnitude for calorimeters in the future International Large Detector (ILD).
Similarly, due to other design constraints, cells in a realistic calorimeter are not arranged in a regular grid but in more complex geometric patterns.

Simulating calorimeters with more than 10k cells using generative models was first attempted in \cite{Belayneh:2019vyx} (65k channels) 
and \cite{Buhmann:2020pmy} (27k channels). 
While these numbers are still much smaller than the entire calorimeter, they allow simulating a slice in $\eta$--$\phi$ large enough to fully contain a shower with realistic granularity. 
Based on such slices, entire calorimeters could e.g. be simulated by conditioning on impact position and angle.

The most accurate generative architecture tested by~\cite{Buhmann:2020pmy} was the co-called bounded information-bottleneck autoencoder (BIB-AE)~\cite{voloshynovskiy2019information}.
It essentially is a VAE with additional GAN-like critic networks.
A key result of this contribution was the correct description of the single hit energy spectrum around the energy deposited by a minimally ionizing particle (MIP).
To this end, an additional post-processing network was trained to fine-tune the output of the generative model.
Further improvement of the fidelity of the generated data was possible by including a secondary density estimation step in the latent space~\cite{Buhmann:2021lxj}, following the Buffer-VAE approach from~\cite{Otten:2019hhl}.

Another challenge lies in simulating the more complex (compared to purely electromagnetic showers) showers initiated by hadrons. 
Here, \cite{Buhmann:2021caf} considered WGAN and BIB-AE architectures for the simulation of positively charged pions in the highly-granular
Analogue Hadron Calorimeter (AHCal).
A potentially important observation was that while the BIB-AE yielded a more accurate initial description of showers, this difference --- at least for energy response and resolution --- largely vanished after processing with standard particle flow reconstruction.
This implies that, depending on the intended downstream use, also simpler generative models might be able to capture relevant characteristics of a shower.

Looking towards the future, a number of challenges remains to be solved:
\begin{itemize}
    \item Simultaneous simulation of different detector geometries and materials for the full depth of a highly granular calorimeter.
    \item Use of non-grid-based architectures (sets, graphs) to capture the geometry of realistic detectors (see e.g. \cite{Kansal:2021cqp})
    \item Multi-dimensional conditioning on energy, impact position, impact angle, and particle type.
    \item Integration in generation workflows of large experimental collaborations (see e.g. \cite{ATL-SOFT-PUB-2018-001} for ATLAS).
    \item Solid treatment of the statistical properties and uncertainties of generated calorimeter data~\cite{Butter:2020qhk,Bieringer:2022cbs}.
\end{itemize}
Nevertheless, the large possible speed-up over alternative methods afforded by generative models makes them a crucial tool in understanding collider data at the highest precision.

\section{Differentiable Programming for Detector Simulation, Design, and Inference}\label{sec:contrib:MK}

%This is an interesting section, but I worry a bit about the scope.  I understood that the three example sections should be vinettes of particular projects.  This section reads as a broad overview of DP for HEP, but it only gives examples from MK.  I am a fan of MK's work, but perhaps it would make more sense to either (1) focus on one particular example (the SHIP example seems most relevant for this WP) and/or (2) be more inclusive in the examples/citations.

%For example, 1907.08209 uses AD with a surrogate for parameter tuning (see Fig. 3).  There is no simulation involved (although there could be), but there is also no simulation involved for madjax (even though it is a very impressive project!).  There are other examples as well.

Differentiable programming for simulation relies on building AD-aware HEP simulation tools. AD uses the chain rule to evaluate derivatives of a function that is represented as a computer program.  AD takes as input program code, whose derivative can be defined, and produces new code for evaluating the program and derivatives. AD typically builds a computational graph, or a directed acyclic graph of mathematical operations applied to an input. Gradients are defined for each operation of the graph, and the total gradient can be evaluated from input to output, called forward mode, and from output to input, called reverse mode or backpropagation in ML.

%Forward mode AD captures the sensitivity of perturbations to a function's inputs on the function's outputs. Forward mode AD is highly efficient for computing the Jacobian of a function's outputs, regardless of the number of outputs, with respect to a single input. Forward derivatives are computed at the same time as the the function itself but, importantly, separate gradient evaluations are needed for derivatives with respect to each input. Reverse mode AD captures the sensitivity of perturbations to a function's outputs on the function's inputs. Reverse mode AD is highly efficient for computing the gradient of one output of a function with respect to the inputs, regardless of the number of inputs. Reverse mode AD is particularly efficient for computing gradients with respect a scalar function, such as a loss function in ML. Reverse mode requires that the function is first evaluated, and then the gradients can be computed recursively backwards starting from the function output until the inputs are reached.

For HEP simulation tools, ideally one would not rewrite the software but instead use AD tools which can merge easily with the existing software. For instance, recent work on \texttt{madjax}~\cite{madjax}, a differentiable matrix element generator, augments python matrix element code generated by \texttt{MadGraph}~\cite{Alwall:2014hca} and merges it with \texttt{JAX}.  

Surrogate models trained to mimic the behavior of high fidelity detector simulators can also be used within DP pipeline. For instance, iteratively trained surrogate models of a \geant\ magnet simulation were used for estimating gradients in a gradient descent optimization  of the magnet system for the SHiP experiment~\cite{Shirobokov:2020tjt}. 
This optimization found more performant and lighter weight designs for a magnetic shield. Similarly, detector surrogates were used to model the smearing induced by detectors on jets, and subsequently to provide gradients for gradient-based unfolding of jet distributions in~\cite{Vandegar:2020yvw}.  Additional applications for fast surrogate models include optimization and design of particle accelerators \cite{edelen:ml2,Koser:ML1,van_der_veken:ml1}, real-time feedback during commissioning and tuning of an accelerator facility \cite{Koser:ML1, van_der_veken:ml1}, and uncertainty quantification of simulated parameters \cite{adelmann:surrogate1,Winklehner:2021qzp}.
These examples show the large potential for such surrogate systems to be used in differentiable inference pipelines for tasks beyond only data generation, see also~\cite{MODE:WP}.

\section{Synergies and a Joint Framework for Detector Simulation}
Different ideas surveyed in this document have shown promising results on individual challenges in detector simulation for HEP. One of the challenges in the future is to identify how different ideas can be combined in a way that benefits the overall scientific community. One of the biggest advantages of \geant\ is the flexibility the software provides, resulting in widespread usage. 

Providing a joint framework for detector simulation supports the testing and benchmarking of new methods as an effective way to promote collaboration between researchers and an ideal environment to keep track of new developments. This direction also streamlines the combination of multiple methods, such as individual detector surrogates, that combined create a full detector simulation.

Data challenges are also an effective method to build collaborations between different scientific communities. Researchers of different backgrounds have the opportunity to discuss and cooperate, promoting new developments. Data challenges are also a good opportunity for transparent comparison of new algorithms. This goal is currently being pursued in ``Fast Calorimeter Simulation Challenge 2022''~\cite{CaloChallenge}.  However, challenges by themselves are not sufficient - resources are required for integrating tools into simulation frameworks (experiment-independent or experiment-specific). 

A joint software framework also opens the possibility for shared development between experiments. Machine learning based models often require large amounts of data for training, restricting the number of users with access to computing centers with available resources for large scale development.\footnote{An alternative solution to this problem in the context of method development and prototyping was recently discussed in~\cite{Sinha:2022ogd}.} However, in a shared software environment, large ML models can be pre-trained in dedicated computing facilities using generic detector geometries. These models can then be later fine-tuned to include experiment-specific information, decreasing the computational burden required to achieve state-of-the-art results. 

One of the biggest challenges of having a unified framework covering multiple experimental facilities is to cope with the differences in computational resources available and experiment-specific software. A possible solution is support for containerized images \cite{merkel2014docker} from experimental collaborations to reproduce their detector simulation routines. This option reduces the need for experiment-specific knowledge while improving software portability. 

Maintenance of the software also becomes crucial. New job positions for HEP software development should also be promoted to ensure future usability and continuity, in order to accelerate future generations of experiments and to ensure that legacy data and results are still accessible. 

\section{Future Directions}\label{sec:future}

The computational complexity required for full detector simulation in high energy physics far exceeds the predicted resources available in future experimental facilities, requiring innovative strategies to accelerate the simulation process while preserving generation quality. Surrogate models are proposed as fast alternatives to replace part of full simulation routines, leveraging advancements in machine learning implemented in heterogeneous computing architectures. 

% Surrogate models presented in this document are based on simplified detectors. Although realistic materials are considered in the studies shown in Secs.~\ref{sec:contrib:EK_WH} and \ref{sec:contrib:CK}, simplified calorimeters usually have a very regular structure. 
While realistic simulations were used in some projects, primarily studies with simplified calorimeters were used to demonstrate the feasibility of new models. These include calorimeter geometries with a very regular structure or with a reduced amount of readout channels, such as the ILD example described in Section~\ref{sec:contrib:EK_WH} or CaloGAN dataset described in Section~\ref{sec:contrib:CK}. However, in a realistic detector this is not usually the case. The number of cells can be large and the geometry irregular. For example, the ATLAS detector calorimeter consists of 173,952 channels of variable size and shape \cite{CERN-LHCC-96-041}, and the CMS High Granularity calorimeter will be constructed using hexagonal wafers \cite{CMS:2017jpq}. Additionally, during a typical shower evolution into the calorimeter only a small portion of the cells ($\mathcal{O}(0.1\%)$) register a signal, leading to a very sparse dataset.
Novel data structures and neural network architectures are required to account for the properties of the data. An example is to represent the calorimeter data in the form of a graph and use a Graph Neural Network to operate on it \cite{Kansal:2020svm,Kansal:2021cqp}. This approach also detaches the method from a particular geometry; data from any type/shape of calorimeter can be converted into a universal graph data structure.

Differential programming can provide powerful new directions in simulator modeling. Building a fully differentiable HEP simulation chain would open a realm of new schemes for optimizing simulations, improving simulation speed, inference and design optimization tasks, and for building physics-informed HEP-ML system that utilize the physics knowledge within HEP simulation software. Dedicated automatic differentiation tools capable of augmenting existing software, rather than requiring complete software rewrites, are needed. New compiler-based source-translation based AD tools, such as \texttt{enzyme}~\cite{enzyme} and \texttt{CLAD}~\cite{clad}, are promising for such tasks.  %Question: do we need HEP-specific tools for converting HEP code to be AD?  Existing and future tools still require some manual input and maybe there should be a dedicated effort to do this?

Applications to realistic scenarios for all ideas will be crucial to identify current limitations and future research directions. While examples in this document have shown promising results, one needs to consider the software environment required to maintain, support, and develop new algorithms. Maintenance of the software is imperative to ensure that algorithms used within experimental collaborations are up to date with the ones available to the wider scientific community.

% In order to have a set of common datasets with which different approaches can be compared to each other, and to further spur development of new approaches, the ``Fast Calorimeter Simulation Challenge 2022'' was created~\cite{CaloChallenge}. 

\Acknowledgements
VM and BN are supported by the U.S. Department of Energy (DOE), Office of Science under contract DE-AC02-05CH11231. MK is supported by the US Department of Energy (DOE) under grant DE-AC02-76SF00515. CK and DS are supported by DOE grant DOE-SC0010008. KP is supported by the Fermi National Accelerator Laboratory, managed and operated by Fermi Research Alliance, LLC under Contract No. DE-AC02-07CH11359 with the U.S. Department of Energy. DW is supported by NSF grant PHY-1912764 and funding from the Heising-Simons Foundation and the Bose Foundation.
Gregor Kasieczka is supported by the Deutsche Forschungsgemeinschaft under Germany’s Excellence Strategy -- EXC 2121 Quantum Universe -- 390833306.

\section{References and bibliography}
\bibliography{bibtex.bib}
\bibliographystyle{unsrturl}

\end{document}

%% file: workshopsymbols.tex
\def\Acknowledgements{\bigskip  \bigskip \begin{center} \begin{large}
             \bf ACKNOWLEDGEMENTS \end{large}\end{center}}

\def\beq{\begin{equation}}
\def\eeq#1{\label{#1}\end{equation}}
\def\eeqn{\end{equation}}

\newenvironment{Eqnarray}%
   {\arraycolsep 0.14em\begin{eqnarray}}{\end{eqnarray}}
\def\beqa{\begin{Eqnarray}}
\def\eeqa#1{\label{#1}\end{Eqnarray}}
\def\eeqan{\end{Eqnarray}}

\let\bar=\overbar

\def\lsim{\mathrel{\raise.3ex\hbox{$<$\kern-.75em\lower1ex\hbox{$\sim$}}}}
\def\gsim{\mathrel{\raise.3ex\hbox{$>$\kern-.75em\lower1ex\hbox{$\sim$}}}}

\def\del{\partial}
\def\Dslash{\not{\hbox{\kern-4pt $D$}}}
\def\dslash{\not{\hbox{\kern-2pt $\del$}}}
\def\pslash{\not{\hbox{\kern-2pt $p$}}}
\def\ETmiss{\not{\hbox{\kern-4pt $E$}}_T}

\def\Dlr{\mathrel{\raise1.5ex\hbox{$\leftrightarrow$\kern-1em\lower1.5ex\hbox{$D$}}}}

\def\MSB{{\bar{M \kern -2pt S}}}
\def\msb{{\bar{\scriptsize M \kern -1pt S}}}

\def\drb{{\bar{\scriptsize D \kern -1pt R}}}

\def\geant{\textsc{Geant4}}
\def\delphes{\textsc{delphes}}